# Properties of Isoscalar Giant Multipole Resonances in medium-mass one-closed-shell nuclei: A semi-microscopic description

M. L. Gorelik[1,*] and M. H. Urin[2,**]

[1] *Moscow Economic School, Moscow 123022, Russia*

[2] *National Research Nuclear University "MEPhI", Moscow 115409, Russia*

*Dedicated to memory of B. A. Tulupov*

## Abstract

The semi-microscopic particle-hole dispersive optical model, having unique abilities in describing main properties of various giant resonances in medium-heavy-mass spherical nuclei, is applied to a number of Isoscalar Giant Multipole Resonances in one-closed-shell nuclei $^{58}$Ni, $^{120}$Sn, and $^{142}$Nd. Some of the calculation results are compared with respective experimental data, including those obtained just recently. Special attention is paid to a comparison of calculated isoscalar multipole strength functions with the respective experimental strength distributions in a wide excitation-energy interval.

## I. INTRODUCTION

Being associated with high-energy collective particle-hole-type nuclear excitations, Giant Resonances (GRs) have been the object of intensive experimental and theoretical studies during many years (see, e. g., Refs. [1, 2]). Of special interest is investigation of Isoscalar Giant Multipole Resonances (ISGMPRs), because a part of these GRs is related to compression modes [1-3]. Being proposed about a decade ago, the semi-microscopic Particle-Hole Dispersive Optical Model (PHDOM) [4] has been successfully implemented to describe main properties of various GRs in medium-heavy-mass doubly-closed-shell nuclei (Refs. [5-7] and references therein). Main properties of an arbitrary GR include the following characteristics defined for a wide excitation-energy interval: the strength distributions, transition densities (these characteristics are related to collective motion), and probabilities (or branching ratios) of direct one-nucleon decay (these characteristics are related to interplay of collective and single-particle motions). Unique abilities of PHDOM in describing the above-listed characteristics are due to combined consideration within the model of the main GR relaxation modes. They are Landau damping, and coupling high-energy (p-h)-type states to the single-particle continuum and to many-quasiparticle configurations (the spreading effect). Within PHDOM, Landau damping and coupling to the single-particle continuum are described microscopically in terms of a partially self-consistent, phenomenological (Woods-Saxon-type) mean field and Landau-Migdal p-h interaction, while the spreading effect is treated in a phenomenological way in terms of the energy-averaged p-h self-energy term. The imaginary part of this term determines the real part via a proper dispersive relationship. Actually, PHDOM is a microscopically-based extension of the standard and non-standard continuum-RPA (cRPA) versions on taking the spreading effect into account (phenomenologically and in average over the excitation energy). The model is a deep modification of the cRPA-based approaches, in which the spreading effect is considered in intuitive ways [8].

---

* mikhailg@mes.ru

** MGUrin@mephi.ru

Some comments regarding the direct use of PHDOM for describing GRs in medium-heavy-mass open-shell spherical nuclei are required. In these nuclei, the effect of nucleon pairing on the "collective" characteristics of "$n\hbar\omega$" multipole GRs ($\hbar\omega$ is the inter-shell energy interval) might be neglected with acceptable accuracy of order $(2\Delta/n\hbar\omega)^2$, where $\Delta$ is the pairing gap. "Smearing" the Fermi surface due to nucleon pairing affects the partial probabilities of GR direct one-nucleon decay. The "smearing" effect (affects the s-p strength distribution) and differences of the experimental and calculated energies of the escaped nucleon (affects the potential-barrier penetrability) might be taken into consideration concerning the probabilities of GR direct one-nucleon decay by means of a recalculating procedure. The comments given above are directly connected with the studies of Refs. [6, 7]. In these studies, the model was implemented to describe various GRs in the one-closed-shell nucleus $^{90}$Zr.

In the present work, following directly to the study of Ref. [6], we implement PHDOM to describing main properties of ISGMPRs, having multipolarity $L = 0 - 3$, and the $L = 0, 2$ overtones of respective GRs in the one-closed-shell nuclei $^{58}$Ni, $^{120}$Sn, and $^{142}$Nd (Sect. II). Bearing in mind the available experimental data for isoscalar multipole strength distributions, including those obtained just recently [9-12], we pay special attention to calculating the respective strength functions (Sect. III). Also, in this Sect., we briefly consider possible extensions of methods employed for analysis of experimental cross sections of GR excitation. Sect. IV contains the summary and our concluding remarks.

## II. PROPERTIES OF ISGMPRs IN $^{58}$Ni, $^{120}$Sn, AND $^{142}$Nd

Since the detailed PHDOM-based description of ISGMPRs in a few medium-heavy-mass spherical nuclei is given in Refs. [5, 6], we restrict the theoretical consideration given below by presenting the model relations and model parameters directly employed in evaluating the characteristics and parameters of GRs to be studied.

Within the model, most GR characteristics can be described, using the effective-field method introduced in nuclear physics by Migdal [13]. In describing the given ISGMPR of multipolarity $L$, the basic quantity is the energy-dependent effective single-particle (s-p) field $\tilde{V}_{LM}(x,\omega) = \tilde{V}_L(r,\omega)Y_{LM}(\vec{n})$ (hereafter, ω is the excitation energy), related to the external field (probing operator) $V_{LM}(x) = V_L(r)Y_{LM}(\vec{n})$. (We neglect the effect of tensor correlations on formation of ISGMPRs ($L > 0$)). The effective field is different from the respective external field due to the core-polarization effect caused by the isoscalar spin-less part of Landau-Migdal p-h interaction. (The strength of this part $F(r)$ is radial dependent). Consequently, the basic equation of the method is the integral equation for the effective-field radial part:

$$\tilde{V}_L(r,\omega) = V_L(r) + \frac{F(r)}{r^2}\int A_L(r,r',\omega)\tilde{V}_L(r',\omega)dr' . \qquad (1)$$

The key quantity in this equation is the $L$-component of the "free" p-h propagator (p-h Green function) $A_L(r,r',\omega)$ described in detail in Refs. [14, 5]. Being defined within PHDOM, this propagator is related to the model of non-interacting independently damping high-energy p-h excitations.

In accordance with the definition employed within PHDOM for the effective field, the latter determines the strength function $S_{V_L}(\omega)$ related to the respective external field $V_{LM}(x)$:

$$S_{V_L}(\omega) = -\frac{1}{\pi} Im \int V_L(r) A_L(r, r', \omega) \tilde{V}_L(r', \omega) dr dr' \, . \quad (2)$$

Being formally considered in infinite excitation-energy interval, the strength function obeys the well-known energy-weighted sum rule (this quantity is defined without the factor $(2L + 1)$):

$$EWSR_{V_L} = \frac{1}{4\pi} \frac{\hbar^2}{2M} A \left\langle \left( \frac{dV_L(r)}{dr} \right)^2 + L(L+1) \left( \frac{V_L(r)}{r} \right)^2 \right\rangle . \quad (3)$$

Here, $A$ is the number of nucleons, $M$ is the nucleon mass, and the averaging $<\ldots>$ is performed over the ground-state nuclear density $n(r) = n^n(r) + n^p(r)$ - the sum of the neutron and proton densities. To compare the strength functions of the given ISGMPR in various nuclei, it is convenient to consider the relative energy-weighted strength function (the $EWSR_{V_L}$ fraction), $y_L(\omega) = \omega S_{V_L}(\omega)/EWSR_{V_L}$. The considered strength functions determine a number of ISGMPR parameters. They are: (i) the GR relative strength $x_L(\omega_{12}) = \int_{\omega_1}^{\omega_2} y_L(\omega) d\omega$ defined for a given excitation-energy interval $\omega_{12} = \omega_1 - \omega_2$; (ii) the GR peak energy $\omega_{L(\text{peak})}$ related to the main maximum of $S_{V_L}(\omega)$; (iii) the GR total width $\Gamma_{L(\text{FWHM})}$ (full width at half maximum (FWHM)) of $S_{V_L}(\omega)$); (iv) the GR energy momentum $m_L^k = \int_{\omega_1}^{\omega_2} \omega^k S_{V_L}(\omega) d\omega$ and ratios of these quantities taken for various values of $k$ and $\omega_{12}$. In particular, we determine the ISGMPR mean energy $\overline{\omega}_L$ as the ratio of respective first to zero momentum both taken for the given $V_L$ and a large excitation-energy interval.

The next GR characteristic related to collective motion is the GR transition density. Strictly speaking, the energy-averaged double transition density $\rho(x, x', \omega) = (rr')^{-2} \sum_{LM} \rho_L(r, r', \omega) Y_{LM}(\vec{n}) Y_{LM}^*(\vec{n}')$, which is determined by the p-h Green function and depends only on nuclear structure [4], should be used in analyzing experimental cross sections of GR excitation. However, in existing computer codes, only one-body ISGMPR transition density is employed (see, e. g., Ref. [1]). For this reason, the projected (i.e., related to the given external field) one-body transition density, $\rho_{V_L}(x,\omega)$ is also considered within PHDOM. This quantity is defined via a convolution of the double transition density with the proper external field [14, 5]. According to the definition, the radial component of the projected transition density, $r^{-2}\rho_{V_L}(r,\omega)$ might be expressed in terms of the radial component of the respective effective field:

$$\frac{1}{r^2}\rho_{V_L}(r,\omega) = -\frac{1}{\pi}\mathrm{Im}\tilde{V}_L(r,\omega)/[F(r)S_{V_L}^{1/2}(\omega)] \,. \quad (4)$$

The projected transition density is normalized by the condition (valid at arbitrary ω): the squared convolution of this density with the related external field equals to the respective strength function.

The possibility to describe direct one-nucleon decay of various GRs belongs to unique abilities of PHDOM. Within the model, the decay amplitude is proportional to the effective-field matrix element taken between proper s-p bound- and continuum-state wave functions. The direct-decay partial strength function, or the differential probability of ISGMPR direct decay accompanied by population of one-hole $\mu^{-1}$ state of the product nucleus, $S_{V_L,\mu}^{\uparrow,\alpha}(\omega)$, is determined by respective squared amplitudes and includes also the squared kinematic factors together with the occupation number of respective s-p level, $n_\mu^\alpha$ ($\alpha = n, p$ is the isobaric index) [15,5]:

$$S_{V_L,\mu}^{\uparrow,\alpha}(\omega) = \sum_{(\lambda)} n_\mu^\alpha (t_{(\lambda)(\mu)}^L)^2 \,|\int \chi_{\varepsilon=\varepsilon_\mu+\omega,(\lambda)}^{\alpha\,*}(r)\tilde{V}_L(r,\omega)\chi_\mu^\alpha(r)dr|^2 \,. \quad (5)$$

Here, $\chi_{\varepsilon>0,(\lambda)}^\alpha(r)$ and $\chi_\mu^\alpha(r)$ are the radial parts of respective one-nucleon continuum-state and bound-state wave functions. The partial branching ratio of ISGMPR direct one-nucleon decay from the excitation-energy interval $\omega_{12}$ into channel $\mu$, $b^{\uparrow,\alpha}_{L,\mu}$, is usually defined as ratio of the direct-decay partial strength function to the full strength function, $S_{V_L}(\omega)$, each integrated over the above-mentioned interval.

$$b^{\uparrow,\alpha}_{L,\mu}(\omega_{12}) = \int_{\omega_{12}} S^{\uparrow,\alpha}_{V_L,\mu}(\omega)d\omega \,/\, \int_{\omega_{12}} S_{V_L}(\omega)d\omega \,. \quad (6)$$

Finally, the partial branching ratios determine the total branching ratios: $b^{\uparrow,\alpha}_{L,tot} = \sum_{\mu} b^{\uparrow,\alpha}_{L,\mu}$ and $b^{\uparrow}_{L,tot} = \sum_{\alpha=n,p} b^{\uparrow,\alpha}_{L,tot}$. Within cRPA (i.e., ignoring the spreading effect), $b^{\uparrow,cRPA}_{L,tot} = 1$ (the unitary condition). Therefore, being compared with unity, the value of $b^{\downarrow}_{L} = 1 - b^{\uparrow}_{L,tot}$ might be considered as an estimation of relative contribution of the spreading effect in formation of the given GR.

Completing the list of model relations employed in the description of $L = 0-3$ ISGMPRs, we consider the choice of the external-field radial part, $V_L(r)$. In describing the real main-tone $L = 2,3$ ISGMPRs (ISGQR, ISGOR), this part and related $EWSR_{V_L}$ of Eq. (3) are the following:

$$V_{L=2,3}(r) = r^L,\ EWSR_{V_{L=2}} = \frac{1}{4\pi}\frac{\hbar^2}{2M} A\ 10\ \langle r^2\rangle,\ EWSR_{V_{L=3}} = \frac{1}{4\pi}\frac{\hbar^2}{2M} A\ 21\ \langle r^4\rangle, \quad (7)$$

where brackets $<\ldots>$ mean averaging over the nucleus ground-state density. It is noteworthy, that from the microscopic point of view, these GRs are not related to the compression modes. The respective transition-density radial dependence $\rho_{V_L}(r, \omega_{L(peak)})$ exhibits node-less behavior. Being related to the compression modes, the "popular" $L = 0,1$ ISGMPRs (ISGMR, ISGDR) are, actually, the first-order overtones of the respective zero-energy spurious states (ss). The $0^+$ spurious state might be associated with the nucleus ground state. The $1^-$ spurious state, which is related to center-of-mass motion, might be described within the model (actually, within cRPA), using the probing-operator radial part $V^{ss}_{L=1} = r$ and the properly adjusted isoscalar spin-less part of Landau-Migdal forces [5]. In fact, this adjusting procedure is related to translation invariance of the model Hamiltonian. In describing ISGMR and ISGDR, the probing-operator radial parts and related $EWSR_{V_L}$ are taken as:

$$V_{L=0,1}(r) = r^L\left(r^2 - \eta_{L=0,1} < r^2 >\right), EWSR_{V_{L=0}} = \frac{1}{4\pi}\frac{\hbar^2}{2M} A\, 4\, \langle r^2 \rangle\ (\eta_{L=0} = 1),$$

$$EWSR_{V_{L=1}} = \frac{1}{4\pi}\frac{\hbar^2}{2M} A\, \langle 11 r^4 - 10\eta_{L=1} < r^2 > r^2 + 3(\eta_{L=1} < r^2 >)^2 \rangle. \quad (8)$$

The parameters $\eta_L$ are found from the condition of absence of the spurious-state excitation by the overtone external field. In describing ISGMR within any unitary model, the constant term in the expression for $V_{L=0}(r)$ might be omitted ($\eta_{L=0} = 0$). The use in this expression of the value $\eta_{L=0} = 1$ within the PHDOM-based description of ISGMR is a result of restoration of model unitarity [15]. Using the method proposed in Ref. [5] for describing the $1^-$ spurious state, one gets the expression for the parameter $\eta_{L=1}$, which leads, within the model, to the value $\eta_{L=1} \approx 5/3$. This value is used below in the PHDOM-based description of ISGDR. Since ISGMR and ISGDR are the first-order overtones, the related transition densities $\rho_{V_L}(r, \omega_{L(peak)})$ exhibit the one-node radial dependence. Among overtones of real ISGMPRs, the overtone of ISGMR (i.e., ISGMR2) and the overtone of ISGQR (i.e., ISGQR2) have the lowest excitation energy. The radial part of the respective external fields and related $EWSR_{V_L}$ are taken as:

$$V^{ov}_{L=0,2}(r) = r^2\left(r^2 - \eta^{ov}_{L=0,2} < r^2 >\right),$$

$$EWSR_{V_{L=0}^{ov}} = \frac{1}{4\pi}\frac{\hbar^2}{2M}A\, 4\langle 4r^6 - 4\eta_{L=0}^{ov} < r^2 > r^4 + (\eta_{L=0}^{ov} < r^2 >)^2 r^2\rangle,$$

$$EWSR_{V_{L=2}^{ov}} = \frac{1}{4\pi}\frac{\hbar^2}{2M}A\, \langle 22r^6 - 28\eta_{L=2}^{ov} < r^2 > r^4 + 10(\eta_{L=2}^{ov} < r^2 >)^2 r^2\rangle \quad (9)$$

The adjusted parameters $\eta_L^{ov}$ are found from the condition of minimal main-tone excitation by the overtone external field: $\int V_L^{ov}(r)\rho_{V_L}\big(r, \omega_{L(peak)}\big)dr = 0$. Actually, ISGMR2 is the second-order overtone and, therefore, the related transition density $\rho_{V_{L=0}^{ov}}(r, \omega_{L=0(peak)}^{ov})$ exhibits the two-node radial dependence.

The following input quantities are used in implementations of PHDOM to the description of ISGMPRs in medium-heavy spherical nuclei: the spin-less part of Landau-Migdal p-h interaction, a realistic phenomenological (of the Woods-Saxon type) partially self-consistent mean field (described in detail in Refs. [16, 5]), and the imaginary part of the properly parameterized energy-averaged p-h self-energy term responsible for the spreading effect (the real part is determined by the imaginary part via the respective dispersive relationship [4]).

(i) The spin-less part of Landau-Migdal forces, $F_{L-M}(x_1, x_2) \to (F(r) + F'\vec{\tau}_1\vec{\tau}_2)\delta(\vec{r}_1 - \vec{r}_2)$, contains the isoscalar and isovector strengths: $F(r) = Cf(r)$ and $F' = Cf'$ ($C = 300$ MeV fm$^3$), respectively. The dimensionless isoscalar strength $f(r)$ is parameterized in accordance with Ref. [13]: $f(r) = f^{ex} + \big(f^{in} - f^{ex}\big)f_{WS}(r)$. The small parameter $f^{in}$ is taken as universal quantity, while the main parameter $f^{ex}$ is slightly varied for each considered nucleus and found from the condition, that the above-discussed $1^-$ spurious state, related to center-of-mass motion and described within cRPA, has near zero excitation-energy and almost totally exhausts the related $EWSR_{V_{L=1}^{ss}}$ [5]. In other words, the mean field determines the above-mentioned strength parameter via translation invariance of the model Hamiltonian.

(ii) The realistic phenomenological, partially self-consistent mean field contains: the central (of the Woods-Saxon type) and spin-orbit isoscalar terms (having intensity $U_0$ and $U_{ls}$, respectively); the isovector term, $\frac{1}{2}\tau^{(3)}v(r)$ ($v(r)$ is the symmetry potential); and the Coulomb potential, $U_C(r)$. The isoscalar terms contain also the Woods-Saxon size and diffuseness parameters ($r_0$ and $a$, respectively). The symmetry potential, $v(r) = 2F'n^{(-)}(r)$ (this relation is a result of the isospin symmetry of the model Hamiltonian), and the Coulomb potential are calculated self-consistently via the neutron-excess density, $n^{(-)}(r) = n^n(r) - n^p(r)$, and the proton density, $n^p(r)$, respectively (neglecting the RPA ground-state correlations). As a result, the strength parameter $f'$ might be related to mean-field parameters. For doubly-closed-shell nuclei $^{48}$Ca, $^{132}$Sn, and $^{208}$Pb, the list of mean-field parameters (deduced from an analysis of single-quasiparticle spectra in the respective even-odd and odd-even nuclei) is given, in particular, in Refs. [5, 6]. Using these data as the basis for the proper interpolation procedure [6], we get the values of mean-field parameters for one-closed-shell nuclei $^{58}$Ni, $^{120}$Sn, and $^{142}$Nd considered below. These values together with values of the strength parameters $f^{ex}$, $f^{in}$ and the external-field parameters $\eta^{ov}_{L=0,2}$, $< r^2 >$ are listed in Table I.

(iii) The strength of the imaginary part of the energy-averaged self-energy term responsible for the spreading effect is parameterized by a three-parametric function of the excitation-energy [5, 6]. In the present study, we use the same (universal) "spreading" parameters, which have been employed in Refs. [5, 6] for the PHDOM-based description of the total width (FWHM) of $L = 0 - 3$ ISGMPRs in the above-mentioned closed-shell nuclei.

We start the presentation of the results obtained within this study from the strength functions $S_{V_L}(\omega)$ and relative energy-weighted strength functions $y_L(\omega)$ evaluated within PHDOM for $L = 0 - 3$ ISGMPRs in $^{58}$Ni, $^{120}$Sn, and $^{142}$Nd. The relative energy-weighted strength functions $y_L(\omega)$, each evaluated for a large excitation-energy interval (given in Tables II - IV), are shown in Figs. 1-3. Being deduced from the calculated strength function $S_{V_L}(\omega)$ of the considered GRs, the relative strength $x_L$and the mean energy $\bar{\omega}_L$, evaluated for the excitation-energy interval $\omega_{12}$, the peak energy $\omega_{L(\mathrm{peak})}$ and the total width $\Gamma_{L(FWHM)}$ are given in Tables II - IV together with available experimental data. In applying to ISGDR, ISGOR, and ISGMR2, ISGQR2 the above-listed parameters are given separately for the low-energy (LE) and high-energy (HE) components of these GRs. Appearance of LE-component of the strength functions of "$n\hbar\omega$" GRs ($n \geq 3$) is a result of a contribution of "$(n-2)\hbar\omega$" p-h configurations in formation of these strength functions.

The next GR characteristic related to collective motion (and to the chosen external field) is the projected transition density, having the (one-dimensional) radial part, $\rho_{V_L}(r,\omega)$. These parts, evaluated within PHDOM at $\omega = \omega_{L(peak)}$ for $L = 0 - 3$ ISGMPRs in nuclei under consideration, are shown in Figs. 4 - 6. The transition densities, calculated for the monopole and quadrupole GRs at their peak energy, are further used to evaluate (accordingly to the above-given prescription) the external-field parameters $\eta^{ov}_{L=0,2}$. Obtained for nuclei under consideration, the values of these parameters are given in Table I together with the values $< r^2 >$. The proper external fields are used to evaluate within PHDOM the characteristics and parameters of ISGMR2 and ISGQR2. The relative energy-weighted strength functions and radial transition densities (taken at the GR peak energy) calculated for these GRs in nuclei under consideration are shown in Figs. 1, 2 and Figs. 4, 5 respectively. The calculated overtone-GR parameters are given in Tables II - IV. In Tables III, IV the energies of relatively weak maxima of the strength-function HE-component of ISGMR2 are given in brackets.

Evaluated within PHDOM the main characteristics and parameters of considered $L = 0 - 3$ ISGMPRs include also partial probabilities of direct one-nucleon decay, $S^{\uparrow,\alpha}_{V_L,\mu}(\omega)$, and calculated on this basis the partial (and total) branching ratios, $b^{\uparrow,\alpha}_{L,\mu}$ (and $b^{\uparrow,\alpha}_{L,tot}$), of mentioned decay. For nuclei under consideration, the calculated total branching ratios are presented for both subsystems, whereas the partial branching ratios are given only for the closed-shell subsystem (Tables V – VII). The excitation-energy intervals $\omega_1 - \omega_2$ employed in calculating the branching ratios are also given in these Tables together with the respective fraction parameters $x_L$. A comparison of the calculated partial branching ratio with the respective experimental value, when it becomes available, might be performed with the use of a recalculating procedure described in Ref. [7].

## III. ISOSCALAR MULTIPOLE STRENGTH DISTRIBUTIONS

In recent years, the strength distributions of ISGMPRs in a few medium-heavy-mass spherical nuclei were deduced from inelastic $\alpha$-scattering cross sections measured in large excitation-energy intervals [9 - 12]. In this Section, we compare these experimental strength distributions with the respective strength functions evaluated within PHDOM in this study and in previous studies of Refs. [5, 6]. The experimental and evaluated strength functions of ISGMR in $^{48}$Ca, $^{58}$Ni, $^{90}$Zr, $^{120}$Sn, $^{142}$Nd, $^{208}$Pb (Fig. 7) and of ISGDR in $^{90}$Zr [6] are in a reasonable agreement. However, the experimental and evaluated strength functions of ISGQR in $^{90}$Zr [6] and $^{142}$Nd (Fig. 8) are in clear disagreement. As noted in Ref. [9], the total relative strength ISGQR in $^{142}$Nd, $x_{L=2}$, essentially exceeds 100%. For this reason the experimental quadrupole strength distribution might be also related to ISGQR2. Main characteristics of ISGQR2 in $^{90}$Zr and $^{142}$Nd are evaluated within PHDOM (Ref. [6] and this study (Sect. II), respectively). The evaluated ISGQR2 strength functions are shown in Fig. 9, while the ISGQR2 transition density (taken at the GR peak energy) is shown in Fig. 5 ($^{142}$Nd). Possibly, the above-mentioned disagreement is explained by another choice (as compared with PHDOM) of the overtone transition density and probing operator employed in Ref. [9] to deduce the ISGQR2 strength distribution from experimental inelastic $\alpha$-scattering cross sections. Problems also appear in describing experimental ISGDR and ISGOR strength functions presented in the second Ref. [9] for $^{142}$Nd. Since the radial parts of isoscalar multipole operators $V_L(r)$ employed in Ref. [9] are different from respective quantities of Eqs. (7)-(9), we compare in Fig. 10 the experimental and calculated normalized strength functions $s_{V_L}(\omega)$ defined as follows:

$$s_{V_L}(\omega) = S_{V_L}(\omega) / \int S_{V_L}(\omega) d\omega \ . \qquad (10)$$

As in case of ISGQR, we note the marked excess above 100% for the experimental total relative strength of ISGDR and ISGOR in $^{142}$Nd (Table IV).

These puzzles and, also, the request of experimentalists to get theoretical estimations of strength functions with inclusion of distant GR tails (see, e.g., Ref. [11]) stimulate us to propose within PHDOM an extension of the DWBA-based method employed to deduce the strength functions from experimental reaction cross sections of ISGMPR excitation. Following Ref. [14], we start from the general statement, that in spherical nuclei the ISGMPR strength function, related to a given s-p probing operator (having the radial part, $V_L(r)$), is determined by the double spatial convolution of the $L$-component of the double transition density (having the radial part $(rr')^{-2}\rho_L(r,r',\omega)$) with the probing operator:

$$S_{V_L}(\omega) = \int V_L(r)\rho_L(r,r',\omega)V_L(r')drdr'. \quad (11)$$

Within the collective model (CM), any ISGMPR of multipolarity $L$ is considered as a single-level excitation characterized by: the energy $\omega_L^{CM}$, $\delta$-function energy dependence of the strength function $S_{V_L}^{CM}(\omega)$, and one-body (energy-independent) transition density, having the radial part $r^{-2}\rho_{V_L}^{CM}(r)$. The CM transition density is normalized to the respective sum rule: $\omega_L^{CM}\left(\int V_L(r)\rho_{V_L}^{CM}(r)dr\right)^2 = EWSR_{V_L}$. Within CM, the radial part of the double transition density is expressed in terms of one-body transition densities as follows:

$$\rho_L^{CM}(r,r',\omega) = \rho_{V_L}^{CM}(r)\rho_{V_L}^{CM}(r')S_{V_L}^{CM}(\omega)\omega_L^{CM}/EWSR_{V_L}. \quad (12)$$

To adopt the collective model to the DWBA-based description of experimental cross sections, the CM double transition density of Eq. (12) is "smeared" by the substitution: $\omega_L^{CM}S_{V_L}^{CM}(\omega) \to \omega S_{V_L}^{expt}(\omega)$. As a result, the adopted CM double transition density $\rho_L^{CM,adopt}(r,r',\omega)$ is factorized in terms of the one-body energy-dependent CM transition densities,

$$\rho_{V_L}^{CM,adopt}(r,\omega) = \rho_{V_L}^{CM}(r)\left(\omega S_{V_L}^{expt}(\omega)/EWSR_{V_L}\right)^{1/2}. \qquad (13)$$

These transition densities are actually employed to deduce the strength functions from experimental reaction cross sections of ISGMPR excitation (see, e.g., Ref. [9]).

Within PHDOM, the double transition density $\rho_L(r,r',\omega)$ is approximately factorized in terms of the projected energy-dependent one-body transition densities $\rho_{V_L}(r,\omega)$ (Sect. II). It might be interesting to employ these densities (instead of the CM transition densities of Eq. (13)) within the DWBA-based method of description of inelastic α-scattering cross sections of ISGMPR excitation. The result of such description might be compared with the respective experimental cross sections.

## IV. SUMMARY AND CONCLUSIVE REMARKS

In this study, results of the rather full theoretical investigations of six Isoscalar Giant Multipole Resonances in a few medium-mass one-closed-shell nuclei are presented. Following previous studies of Refs. [5, 6], we implement the semi-microscopic Particle-Hole Dispersive Optical Model to describe the strength functions, transition densities, and branching ratios of direct one-nucleon decay of $L = 0 - 3$ ISGMPRs, including $L = 0, 2$, overtones in the $^{58}$Ni, $^{120}$Sn, and $^{142}$Nd nuclei. The calculated strength functions and GR parameters deduced from these strength functions are compared with available experimental data for nuclei under consideration. The strength functions studied previously for the $^{48}$Ca, $^{90}$Zr and $^{208}$Pb nuclei are also compared with the data. An acceptable agreement is obtained for the strength functions of ISGMR. The agreement might possibly be improved provided that the choice of "spreading" parameters will be specified. Such an attempt has been recently undertaken in the study of Ref. [21], where a detailed PHDOM-based description of experimental Gamow-Teller strength distributions has been obtained for medium-heavy-mass doubly-closed-shell parent nuclei.

As for the isoscalar quadrupole strength distribution, we note the first-time observation of ISGQR2 [9]. In the study of Ref. [9], ISGQR2 in $^{142}$Nd is, actually, attributed to the high-energy component of the ISGQR strength distribution. It means, that the transition density and probing operator employed within the DWBA-based method of describing inelastic $\alpha$-scattering cross sections are taken the same for ISGQR and ISGQR2. Within PHDOM, the related quantities are essentially different (Sects. II, III). For this reason, a further analysis of respective experimental data seems necessary. A similar statement might be also related to the strength functions of ISGDR and ISGOR in $^{142}$Nd in view of the marked difference between the respective experimental and calculated normalized strength functions of the mentioned GRs (Sect. III).

These puzzles and possibilities to extend the DWBA-based method mentioned above by taking into consideration the giant-resonance transition densities evaluated within microscopic (or semi-microscopic) theoretical models stimulate us to propose employing the related transition densities calculated within PHDOM with the use of proper probing operators (Sects. II, III).

**ACNOWLEDJEMENTS**

The authors are thankful to Prof. M.N. Harakeh for interesting discussions, viewing the manuscript and many valuable remarks. The authors thank A. Bahini and S. Bagchi for providing the experimental data shown in the figures. This work is partially supported by the Program "Priority-2030" for the National Research Nuclear University "MEPhI" and Project No. FSWU-2020-0035 of the Ministry of Science and Higher Education of the Russian Federation (M.H.U.).

## REFERENCES

[1] M.N. Harakeh, A. van der Woude, Giant Resonances: Fundamental High-Frequency Modes of Nuclear Excitation, Oxford Science Publications, 2001.

[2] S. Shlomo and V. M. Kolomietz, *Mean Field Theory* (World Scientific, Singapore, 2020).

[3] U. Garg and G. Colò, Progr. Part. Nucl. Phys. **101**, 55 (2018).

[4] M. H. Urin, Phys. Rev. C **87**, 044330 (2013).

[5] M. L. Gorelik, S. Shlomo, B. A. Tulupov, and M. H. Urin, Phys. Rev. C **103**, 034302 (2021).

[6] M. L. Gorelik, S. Shlomo, B. A. Tulupov, and M. H. Urin, Phys. Rev. C **108**, 014328 (2023).

[7] V. I. Bondarenko and M. H. Urin, Phys. Rev. C **106**, 024331 (2022), Phys. Rev. C **109**, 064610 (2024).

[8] M. L. Gorelik, I. V. Safonov, and M. H. Urin, Phys. Rev. C **69**, 054322 (2004); M. H. Urin, Phys. At. Nucl. **74**, 1189 (2011).

[9] M. Abdullah, S. Bagchi, M.N. Harakeh, H. Akimune, D. Das, T. Doi, L.M. Donaldson, Y. Fujikawa, M. Fujiwara, T. Furuno, U. Garg, Y.K. Gupta, K.B. Howard, Y. Hijikata, K. Inaba, S. Ishida, M. Itoh, N. Kalantar-Nayestanaki, D. Kar, T. Kawabata, S. Kawashima, K. Khokhar, K. Kitamura, N. Kobayashi, Y. Matsuda, A. Nakagawa, S. Nakamura, K. Nosaka, S. Okamoto, S. Ota, S. Pal, R. Pramanik, S. Roy, S. Weyhmiller, Z. Yang, J.C. Zamora, Phys. Lett. B **855**, 138852 (2024); Phys. Rev. C **112**, 044316 (2025).

[10] A. Bahini, R. Neveling, P. von Neumann-Cosel, J. Carter, I. T. Usman, P. Adsley, N. Botha, J. W. Brümmer, L. M. Donaldson, S. Jongile, T. C. Khumalo, M. B. Latif, K. C. W. Li, P. Z. Mabika, P. T. Molema, C. S. Moodley, S. D. Olorunfunmi, P. Papka, L. Pellegri, B. Rebeiro, E. Sideras-Haddad, F. D. Smit, S. Triambak, M. Wiedeking, and J. J. van Zyl, Phys. Rev. C **107**, 034312 (2023).

[11] S. D. Olorunfunmi, R. Neveling, J. Carter, P. von Neumann-Cosel, I. T. Usman, P. Adsley, A. Bahini, L. P. L. Baloyi, J. W. Brümmer, L. M. Donaldson, H. Jivan, N. Y. Kheswa, K. C. W. Li, D. J. Marín-Lámbarri, P. T. Molema, C. S. Moodley, G. G. O'Neill, P. Papka, L. Pellegri, V. Pesudo, E. Sideras-Haddad, F. D. Smit, G. F.

Steyn, A. A. Avaa, F. Diel, F. Dunkel, P. Jones, and V. Karayonchev, Phys. Rev. C **105**, 054319 (2022).

[12] Y. K. Gupta, K. B. Howard, U. Garg, J. T. Matta, M. Şenyiğit, M. Itoh, S. Ando, T. Aoki, A. Uchiyama, S. Adachi, M. Fujiwara, C. Iwamoto, A. Tamii, H. Akimune, C. Kadono, Y. Matsuda, T. Nakahara, T. Furuno, T. Kawabata, M. Tsumura, M. N. Harakeh, and N. Kalantar-Nayestanaki, Phys. Rev. C **97**, 064323 (2018).

[13] A. B. Migdal, Theory of Finite Fermi Systems and Applications to Atomic Nuclei (Interscience, New York, 1967).

[14] M. L. Gorelik, S. Shlomo, B. A. Tulupov, and M. H. Urin, Nucl. Phys. A **955**, 116 (2016).

[15] M. L. Gorelik, S. Shlomo, B. A. Tulupov, and M. H. Urin, Nucl. Phys. A **970**, 353 (2018).

[16] G. V. Kolomiytsev, S. Y. Igashov, and M. H. Urin, Phys. At. Nucl. **77**, 1105 (2014).

[17] J. Button, Y.-W. Lui, D.H. Youngblood, X. Chen, G. Bonasera, S. Shlomo, Phys. Rev. C **94**, 034315 (2016).

[18] B.K. Nayak, U. Garg, M. Hedden, M. Koss, T. Li, Y. Liu, P. Madhusudhana Rao, S. Zhu, M. Itoh, and H. Sakaguchi, et al., Phys. Lett. B **637**, 43 (2006).

[19] J. Arroyo, U. Garg, H. Akimune, G. P. A. Berg, D. C. Cuong, M. Fujiwara, M. N. Harakeh, M. Itoh, T. Kawabata, K. Kawase, J. T. Matta, D. Patel, M. Uchida, M. Yosoi, Phys. Rec. C **111**, 014308 (2025)

[20] T. Li, U. Garg, Y. Liu, R. Marks, B.K. Nayak, P.V. Madhusudhana Rao, M. Fujiwara, H. Hashimoto, K. Nakanishi, S. Okumura, M. Yosoi, M. Ichikaa, M. Itoh, R. Matsuo, T. Terazono, M. Uchida, Y. Iwao, T. Kawabata, T. Murakami, H. Sakaguchi, S. Terashima, Y. Yasuda, J. Zenihiro, H. Akimune, K. Kawase, M.N. Harakeh, Phys. Rev. C **81** 034309 (2010).

[21] V. I. Bondarenko and M. H. Urin, LXXV International Conference "NUCLEUS-2025" (Saint Petersburg, Russia, 1-6 July 2025), Book of abstracts p.27, ISBN: 978-5-86763-496-4.

**TABLE I.** The values of mean-field parameters, parameters of the spin-less part of Landau-Migdal forces, and parameters of isoscalar external fields, all employed in PHDOM-based calculations of characteristics of ISGMPRs in nuclei under consideration. (Notations are given in the text). The values $r_0 = 1.21$ fm and $f^{in} = 0.0875$ are taken as universal quantities.

| Nucleus | $U_0$, MeV | $U_{ls}$, MeV fm$^2$ | $a$, fm | $f'$ | $-f^{ex}$ | $\eta^{ov}_{L=0}$ | $\eta^{ov}_{L=2}$ | $< r^2 >$, fm$^2$ |
|---|---|---|---|---|---|---|---|---|
| $^{58}$Ni | 54.49 | 32.93 | 0.586 | 1.11 | 2.564 | 3.11 | 1.80 | 13.05 |
| $^{120}$Sn | 55.38 | 35.86 | 0.629 | 1.01 | 2.610 | 2.71 | 1.81 | 20.52 |
| $^{142}$Nd | 55.66 | 35.97 | 0.636 | 0.991 | 2.568 | 2.63 | 1.76 | 22.70 |

**TABLE II.** The parameters of ISGMPRs in $^{58}$Ni evaluated within PHDOM. (Notations are given in the text). Available experimental data are also given.

| $L$ | $\omega_1 - \omega_2$, MeV | $x_L$, % | $\bar{\omega}_L$, MeV | $\omega_{L(\text{peak})}$, MeV | $\Gamma_{L(\text{FWHM})}$, MeV |
|---|---|---|---|---|---|
| 0 | 10 – 30 | 97 | 19.5 | 19.0 | 4.1 |
| expt. [10] | 10 – 24.5 | - | 18.31$\pm$0.23 | 17.8$\pm$0.4 | 5.4$\pm$0.4 |
| expt. [17] | 10 – 35 | $82^{+11}_{-9}$ | $19.20^{+0.44}_{-0.19}$ | 18.43$\pm$0.15 | 7.41$\pm$0.13 |
| expt. [18] | 10.5 – 32.5 | $92^{+3}_{-4}$ | $19.9^{+0.7}_{-0.8}$ | - | - |
| expt. [19] | 10.5 – 28.5 | 79$\pm$2 | 17.8$\pm$0.1 | 17.5$\pm$0.1 | 4.6$\pm$0.1 |
| 0 $ov$ (LE) | 10 – 27 | 30 | 19.9 | 13.0, 22.0, 23.7 | - |
| 0 $ov$ (HE) | 27 – 50 | 61 | 36.1 | 33.0 | - |
| 1 (LE) | 8 – 18 | 14 | 12.5 | 9.3, 12.8 | 0.8, 3.0 |
| expt. [17] | - | 4$\pm$2 | 17.42$\pm$0.25 | - | $3.94^{+0.36}_{-0.34}$ |
| 1 (HE) | 18 – 38 | 78 | 28.4 | 29.2 | 13.2 |
| expt. [17] | - | 86$\pm$12 | 34.06$\pm$0.30 | - | $19.52^{+0.41}_{-0.40}$ |
| 2 | 10 – 20 | 71 | 15.2 | 15.4 | 2.5 |
| | 20 – 30 | 14 | 23.4 | 24.6 | - |
| 2 $ov$ (LE) | 10 – 27 | 39 | 19.0 | 19.7 | - |
| 2 $ov$ (HE) | 27 – 50 | 56 | 36.3 | 34.6 | - |
| 3 (LE) | 5 – 10 | 15 | 7.3 | 7.1 | 1.0 |
| | 10 – 20 | 14 | 14.3 | - | - |
| 3 (HE) | 20 – 35 | 54 | 27.6 | 27.6 | 3.2 |

**TABLE III.** The same as in Table II but for $^{120}$Sn.

| $L$ | $\omega_1 - \omega_2$, MeV | $x_L$, % | $\overline{\omega}_L$, MeV | $\omega_{L(\mathrm{peak})}$, MeV | $\Gamma_{L(\mathrm{FWHM})}$, MeV |
|---|---|---|---|---|---|
| 0 | 7 – 25 | 97 | 16.2 | 16.3 | 4.0 |
| expt. [10] | 10 – 24.5 | - | 16.24±0.39 | 15.5±0.4 | 5.6±0.4 |
| expt. [20] | 10.5 – 20.5 | 108 ±7 | 15.7±0.1 | 15.4±0.2 | 4.9±0.5 |
| 0 $ov$ (LE) | 7 – 24 | 35 | 15.3 | 9.1, 12.1, 18.4, 20.3 | - |
| 0 $ov$ (HE) | 24 – 50 | 62 | 33.5 | 27.7, (36.5) | - |
| 1 (LE) | 5 – 15 | 14 | 10.1 | 6.6, 7.5, 8.8, 11.0, 13.5 | - |
| expt. [20] | - | - | - | 14.7±0.1 | 5.9±0.3 |
| 1 (HE) | 15 – 35 | 78 | 25.2 | 27.6 | 8.8 |
| expt. [20] | - | 150±3 | - | 26.0±0.4 | 13.1±1.9 |
| 2 | 7 – 17 | 72 | 12.1 | 12.4 | 2.7 |
| | 17 – 25 | 14 | 20.2 | 20.0 | - |
| 2 $ov$ (LE) | 7 – 27 | 50 | 16.4 | 13.8, 19.6 | - |
| 2 $ov$ (HE) | 27 – 50 | 50 | 35.0 | 35.3 | - |
| 3 (LE) | 3 – 15 | 29 | 6.6 | 4.6, 5.9, 8.7 | - |
| 3 (HE) | 15 – 35 | 63 | 23.1 | 22.6 | 3.6 |

**TABLE IV.** The same as in Table II but for $^{142}$Nd. The respective experimental values (given with errors) are taken from Ref. [9].

| $L$ | $\omega_1 - \omega_2$, MeV | $x_L$, % | $\overline{\omega}_L$, MeV | $\omega_{L(\text{peak})}$, MeV | $\Gamma_{L(\text{FWHM})}$, MeV |
|---|---|---|---|---|---|
| 0 | 10 – 30 | 99 | 16.2 | 16.1 | 3.9 |
| | 10 – 22 | 93 | 15.8 | | |
| | 10 – 22 | $103.9^{+10.9}_{-14.3}$ | $15.3 \pm 0.1$ | - | $3.3 \pm 0.2$ |
| 0 $ov$ (LE) | 10 – 22 | 25 | 16.4 | 12.5, 14.5, 18.7 | - |
| 0 $ov$ (HE) | 22 – 50 | 74 | 32.5 | 29.6, (34.2), (36.3) | - |
| 1 (LE) | 5 – 16 | 13 | 10.9 | 7.2, 8.3, 9.7, 12.2 | - |
| | 10 – 16 | 10 | 12.6 | | |
| | 10 – 16 | $51.3^{+3.1}_{-3.6}$ | $14.1 \pm 0.3$ | | $6.4 \pm 0.5$ |
| 1 (HE) | 16 – 50 | 82 | 25.9 | 26.5 | 7.7 |
| | 16 – 30 | 66 | 24.4 | | |
| | 16 – 30 | $99.5^{+21.0}_{-20.5}$ | $24.1 \pm 0.4$ | - | $7.6 \pm 1.0$ |
| 2 | 5 – 17 | 73 | 11.5 | 11.9 | 2.6 |
| | 17 – 30 | 17 | 20.9 | 19.3 | - |
| | 10 – 16 | 60 | 12.3 | 11.9 | 2.6 |
| | 10 – 16 | $109.6^{+3.5}_{-3.5}$ | $12.4 \pm 0.1$ | - | $6.3 \pm 0.2$ |
| 2 $ov$ (LE) | 5 – 25 | 41 | 16.7 | 12.5, 14.8, 19.6 | - |
| 2 $ov$ (HE) | 25 – 50 | 60 | 33.7 | 34.8 | - |
| 3 (LE) | 3 – 17 | 35 | 6.3 | 4.1, 5.6 | - |
| | 10 – 17 | 12 | 13.1 | | |
| | 10 – 17 | $37.6 \pm 5.4$ | $12.7 \pm 0.7$ | - | - |
| 3 (HE) | 17 – 50 | 64 | 23.7 | 21.6 | 3.7 |
| | 17 – 30 | 54 | 22.4 | | |
| | 17 – 30 | $197.0 \pm 13.2$ | $23.2 \pm 1.0$ | - | - |

**TABLE V.** Evaluated within PHDOM for ISGMPRs in [58]Ni the total and proton partial (into the channel $\mu$) branching ratios of direct one-nucleon decay (in percentage). The considered excitation-energy intervals $\omega_1 - \omega_2$ and respective fraction parameters $x_L$ are also given.

| $\mu^{-1}$ | $\varepsilon_\mu$, MeV | $b^{\uparrow}_{L=0,\mu}$ | $b^{\uparrow}_{L=1,\mu}$ | $b^{\uparrow}_{L=2,\mu}$ | $b^{\uparrow}_{L=3,\mu}$ | $b^{\uparrow ov}_{L=0,\mu}$ | $b^{\uparrow ov}_{L=2,\mu}$ |
|---|---|---|---|---|---|---|---|
| $1f_{7/2}$ | -7.15 | 23.2 | 20.1 | 5.5 | 25.2 | 14.6 | 22.3 |
| $1d_{3/2}$ | -12.93 | 5.9 | 7.2 | 0.2 | 2.0 | 9.2 | 6.8 |
| $2s_{1/2}$ | -12.98 | 8.9 | 6.8 | 0.7 | 4.3 | 7.8 | 5.2 |
| $1d_{5/2}$ | -16.93 | 6.7 | 10.1 | - | 4.2 | 11.8 | 9.0 |
| $1p_{1/2}$ | -24.36 | - | 1.8 | - | 0.3 | 2.1 | 1.2 |
| $1p_{3/2}$ | -26.15 | - | 7.7 | - | 0.2 | 3.5 | 2.8 |
| $1s_{1/2}$ | -34.56 | - | - | - | - | 0.1 | 2.2 |
| $b^{\uparrow,p}_{L,tot}$ , % | | 45 | 54 | 6 | 36 | 49 | 49 |
| $b^{\uparrow,n}_{L,tot}$ , % | | 28 | 46 | 10 | 33 | 51 | 49 |
| $\omega_1 - \omega_2$, MeV | | 10 – 30 | 18 – 38 | 10 – 20 | 20 – 35 | 27 – 40 | 27 – 42 |
| $x_L$, % | | 97 | 78 | 71 | 54 | 42 | 44 |

**TABLE VI.** The same as in Table V but for $^{120}$Sn.

| $\mu^{-1}$ | $\varepsilon_{\mu}$, MeV | $b^{\uparrow}_{L=0,\mu}$ | $b^{\uparrow}_{L=1,\mu}$ | $b^{\uparrow}_{L=2,\mu}$ | $b^{\uparrow}_{L=3,\mu}$ | $b^{\uparrow ov}_{L=0,\mu}$ | $b^{\uparrow ov}_{L=2,\mu}$ |
|---|---|---|---|---|---|---|---|
| $1g_{9/2}$ | -10.55 | 1.0 | 6.0 | - | 2.9 | 5.9 | 10.0 |
| $2p_{1/2}$ | -12.46 | 0.7 | 2.9 | - | 0.5 | 3.7 | 2.3 |
| $2p_{3/2}$ | -13.98 | 0.3 | 5.9 | - | 1.2 | 7.2 | 5.1 |
| $1f_{5/2}$ | -14.67 | - | 2.4 | - | 0.2 | 4.5 | 3.1 |
| $1f_{7/2}$ | -18.47 | - | 3.4 | - | - | 5.0 | 5.6 |
| $2s_{1/2}$ | -22.45 | - | 0.5 | - | - | 2.2 | 2.0 |
| $1d_{3/2}$ | -23.72 | - | 0.1 | - | - | 1.5 | 1.1 |
| $1d_{5/2}$ | -25.82 | - | - | - | - | 2.5 | 3.8 |
| $1p_{1/2}$ | -31.61 | - | - | - | - | 0.4 | 0.1 |
| $1p_{3/2}$ | -32.49 | - | - | - | - | 0.3 | 0.1 |
| $b^{\uparrow,p}_{L,tot}$ , % | | 2 | 21 | 0 | 5 | 33 | 33 |
| $b^{\uparrow,n}_{L,tot}$ , % | | 48 | 67 | 16 | 44 | 67 | 65 |
| $\omega_1 - \omega_2$, MeV | | 9 – 23 | 17 – 32 | 8 – 16 | 19 – 26 | 24 – 40 | 28 – 40 |
| $x_L$, % | | 94 | 69 | 68 | 40 | 50 | 38 |

**TABLE VII.** The same as in Table V but for the total and neutron partial branching ratios of direct one-nucleon decay of ISGMPRs in $^{142}$Nd.

| $\mu^{-1}$ | $\varepsilon_\mu$, MeV | $b^{\uparrow}_{L=0,\mu}$ | $b^{\uparrow}_{L=1,\mu}$ | $b^{\uparrow}_{L=2,\mu}$ | $b^{\uparrow}_{L=3,\mu}$ | $b^{\uparrow ov}_{L=0,\mu}$ | $b^{\uparrow ov}_{L=2,\mu}$ |
|---|---|---|---|---|---|---|---|
| $2d_{3/2}$ | -10.79 | 15.0 | 5.6 | 1.3 | 5.4 | 6.6 | 3.2 |
| $3s_{1/2}$ | -10.90 | 6.5 | 2.6 | 2.6 | 2.3 | 4.2 | 1.4 |
| $1h_{11/2}$ | -10.99 | 5.6 | 13.0 | 0.6 | 7.7 | 4.8 | 15.2 |
| $2d_{5/2}$ | -13.31 | 20.2 | 9.9 | 0.4 | 13.9 | 11.0 | 6.2 |
| $1g_{7/2}$ | -13.77 | 2.7 | 6.6 | - | 1.9 | 5.8 | 6.1 |
| $1g_{9/2}$ | -18.80 | 0.05 | 9.2 | - | 1.8 | 6.6 | 10.4 |
| $2p_{1/2}$ | -20.02 | - | 3.6 | - | 0.6 | 3.8 | 2.2 |
| $2p_{3/2}$ | -21.32 | - | 7.8 | - | 0.7 | 7.8 | 5.2 |
| $1f_{5/2}$ | -22.91 | - | 2.9 | - | 0.1 | 4.0 | 3.0 |
| $1f_{7/2}$ | -26.12 | - | 1.2 | - | - | 4.9 | 6.6 |
| $2s_{1/2}$ | -29.21 | - | - | - | - | 2.2 | 2.3 |
| $1d_{3/2}$ | -31.05 | - | - | - | - | 1.6 | 1.4 |
| $1d_{5/2}$ | -32.83 | - | - | - | - | 3.1 | 1.2 |
| $b^{\uparrow,n}_{L,tot}$ , % | | 50 | 62 | 5 | 34 | 67 | 64 |
| $b^{\uparrow,p}_{L,tot}$ , % | | 3 | 23 | 0 | 5 | 33 | 30 |
| $\omega_1 - \omega_2$, MeV | | 11 – 21 | 19 – 31 | 8 – 15 | 18 – 25 | 23 – 38 | 27 – 38 |
| $x_L$, % | | 89 | 65 | 64 | 41 | 53 | 41 |

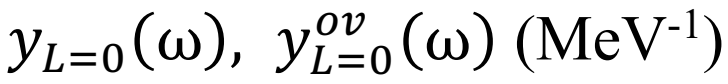

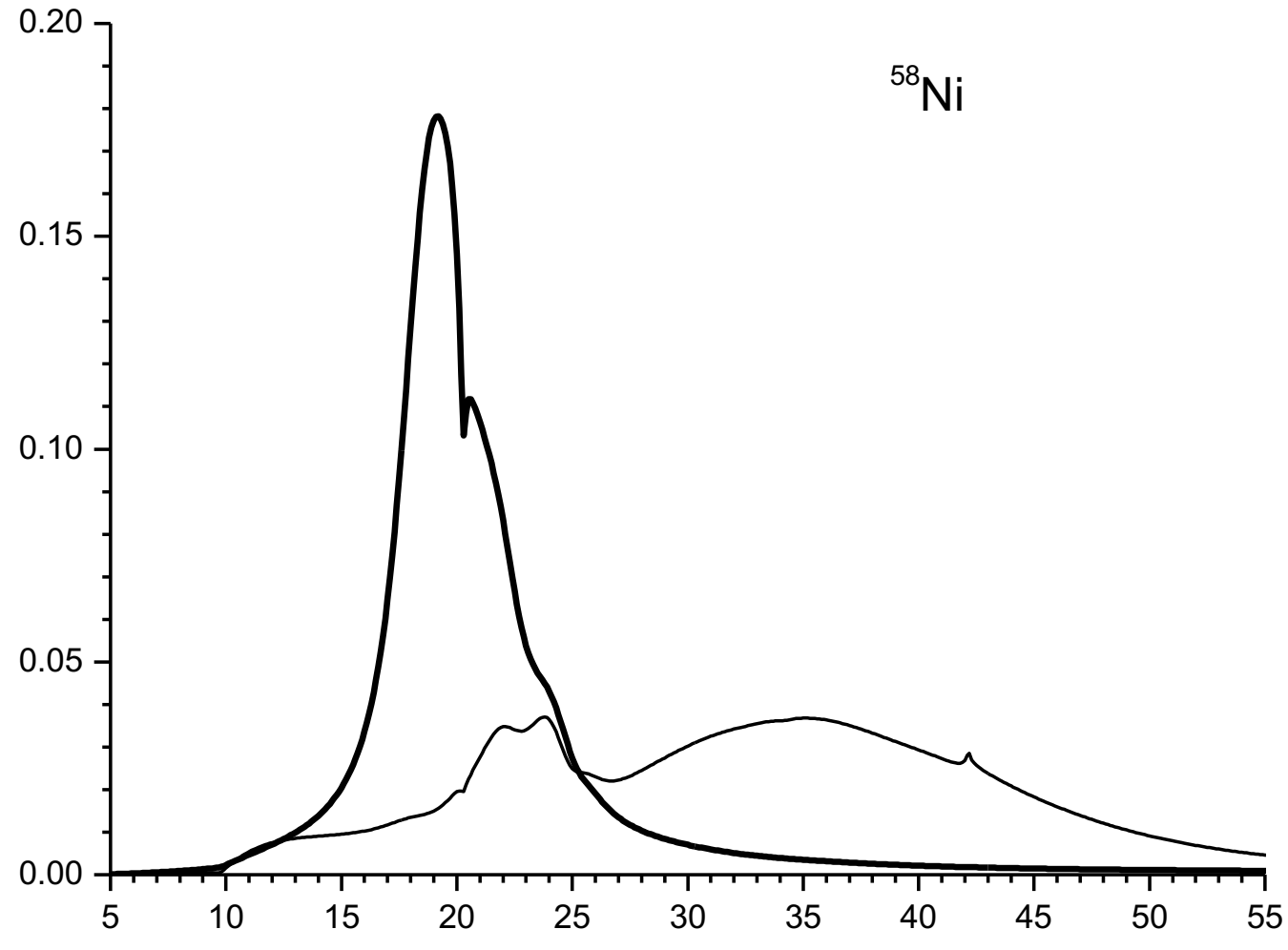

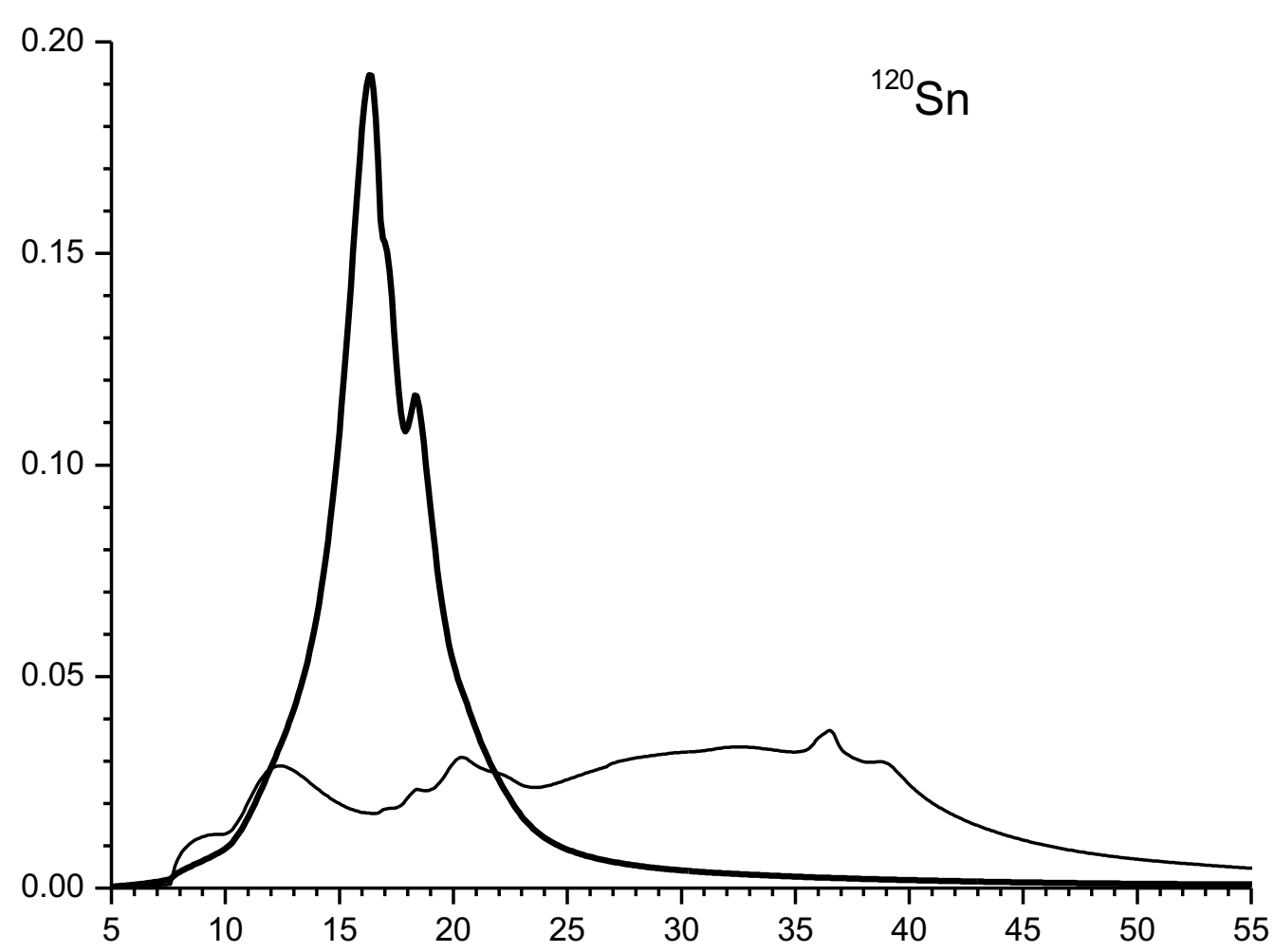

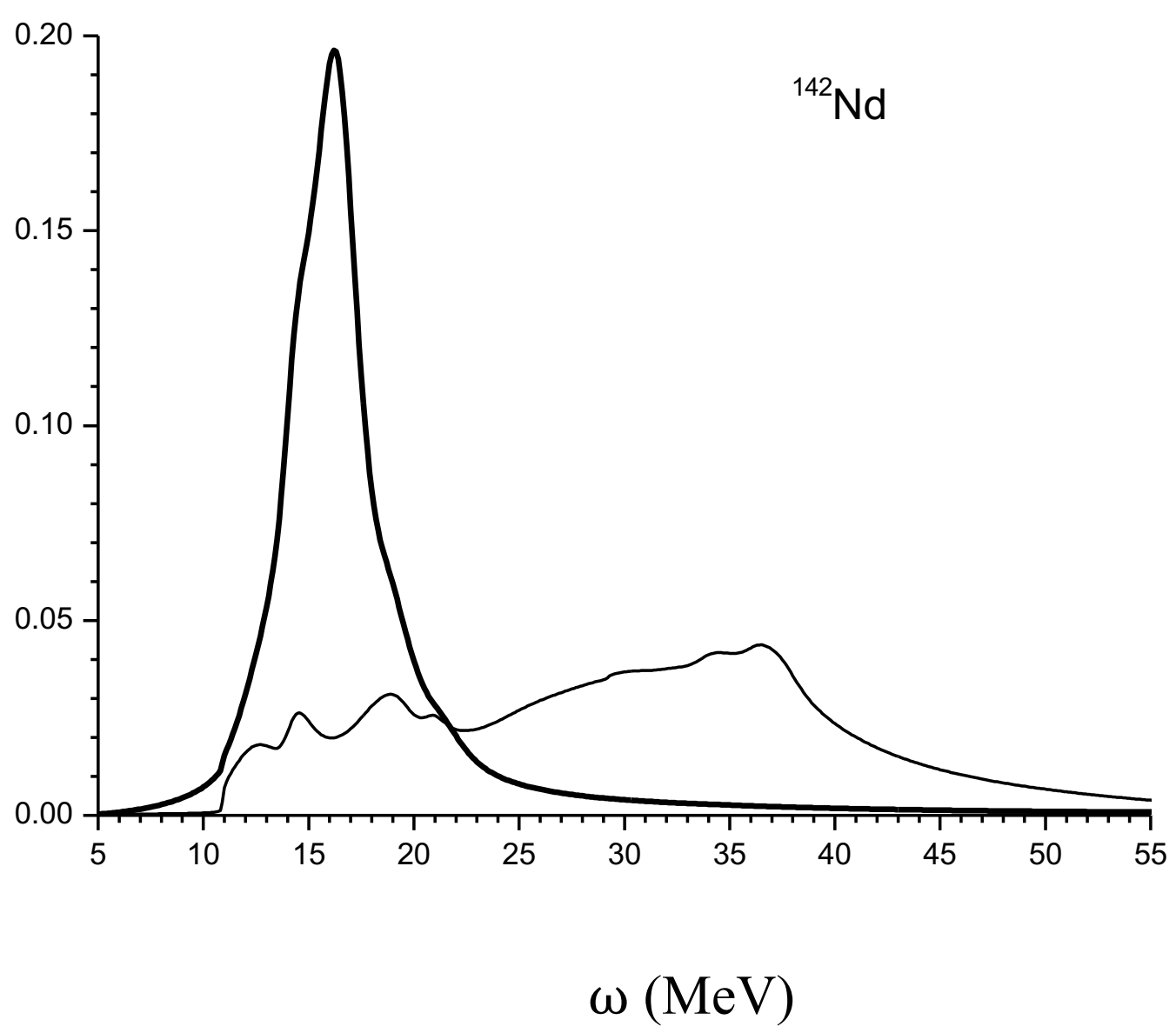

**FIG. 1.** The relative energy-weighted strength functions (EWSR fractions) evaluated within PHDOM for ISGMR (solid line) and ISGMR2 (thin line) in nuclei under consideration.

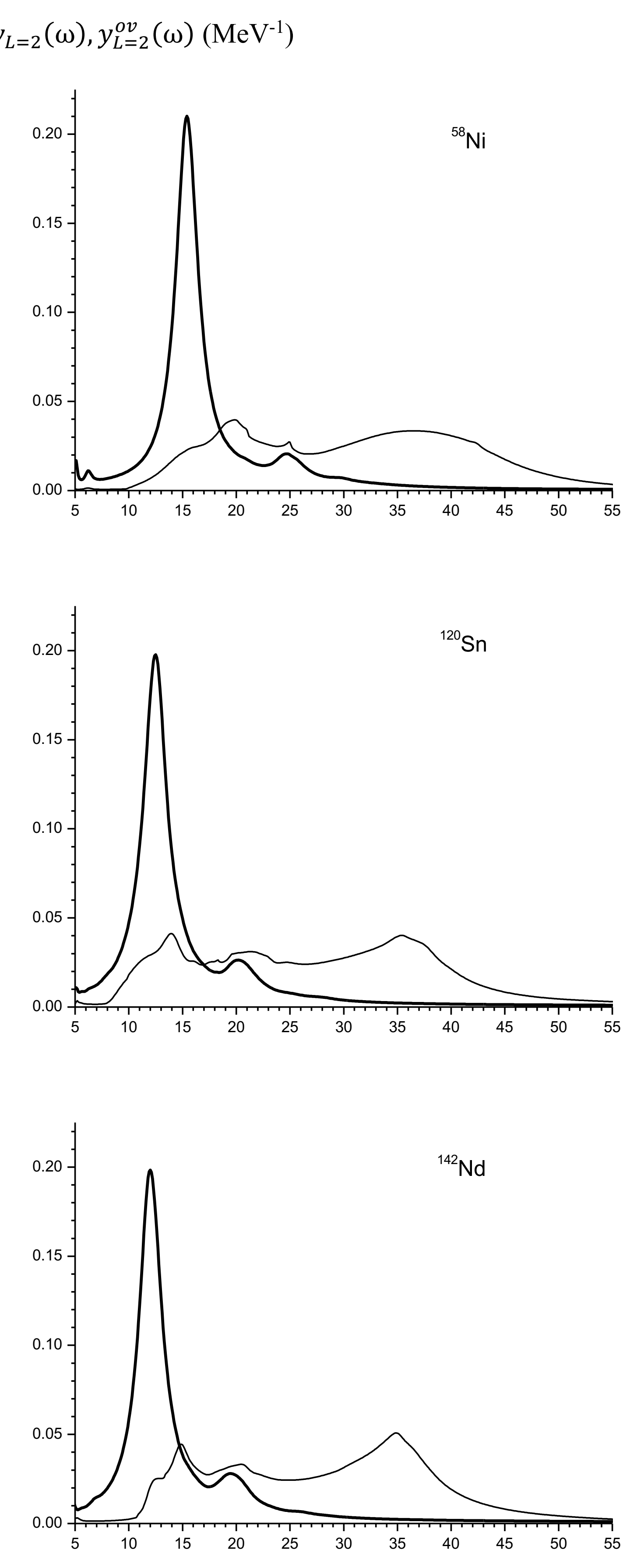

ω (MeV)

**FIG. 2.** The same as in Fig. 1 but for ISGQR (solid line) and ISGQR2 (thin line).

$y_{L=1}(\omega), y_{L=3}(\omega)$ (MeV$^{-1}$)

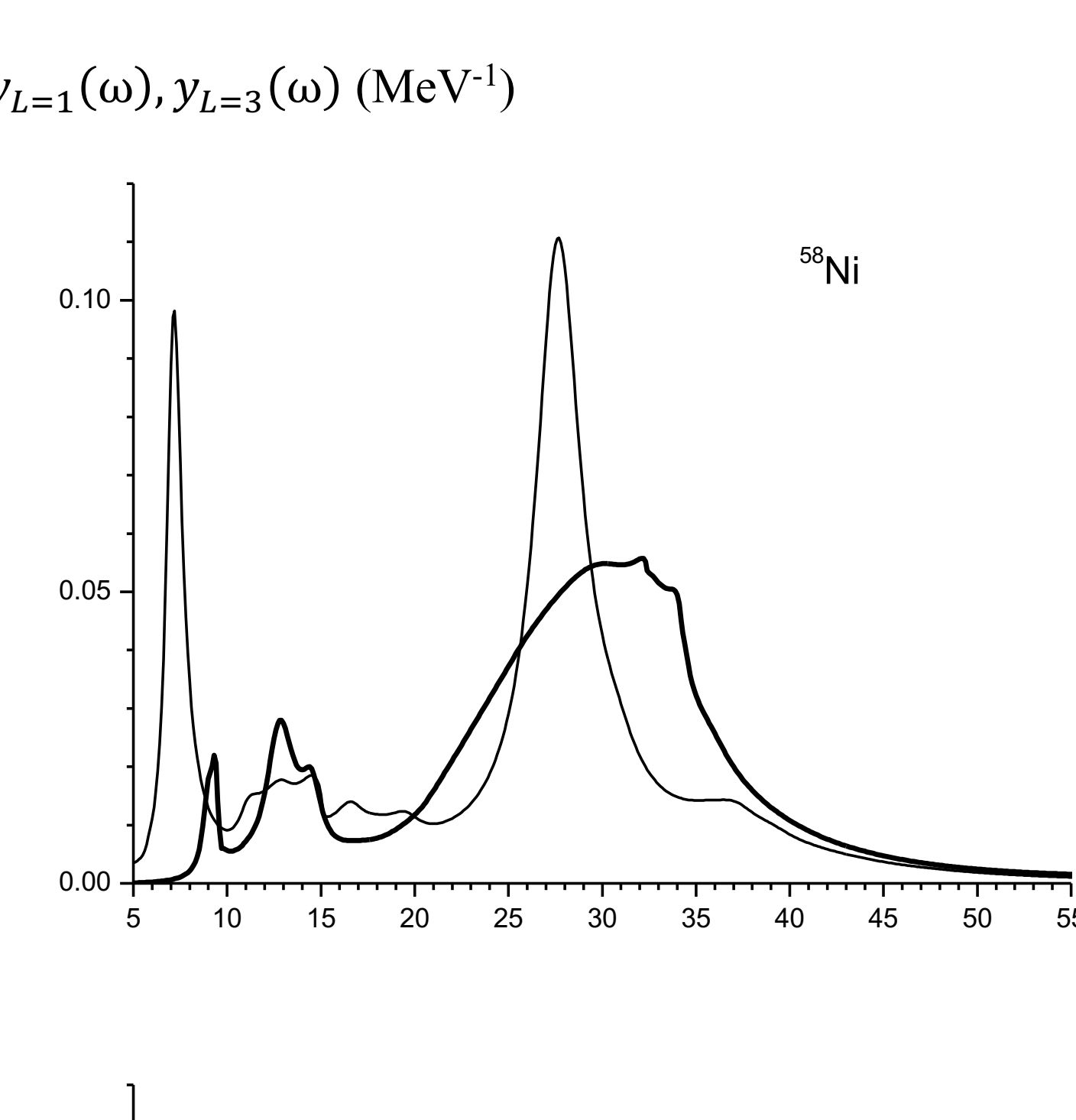

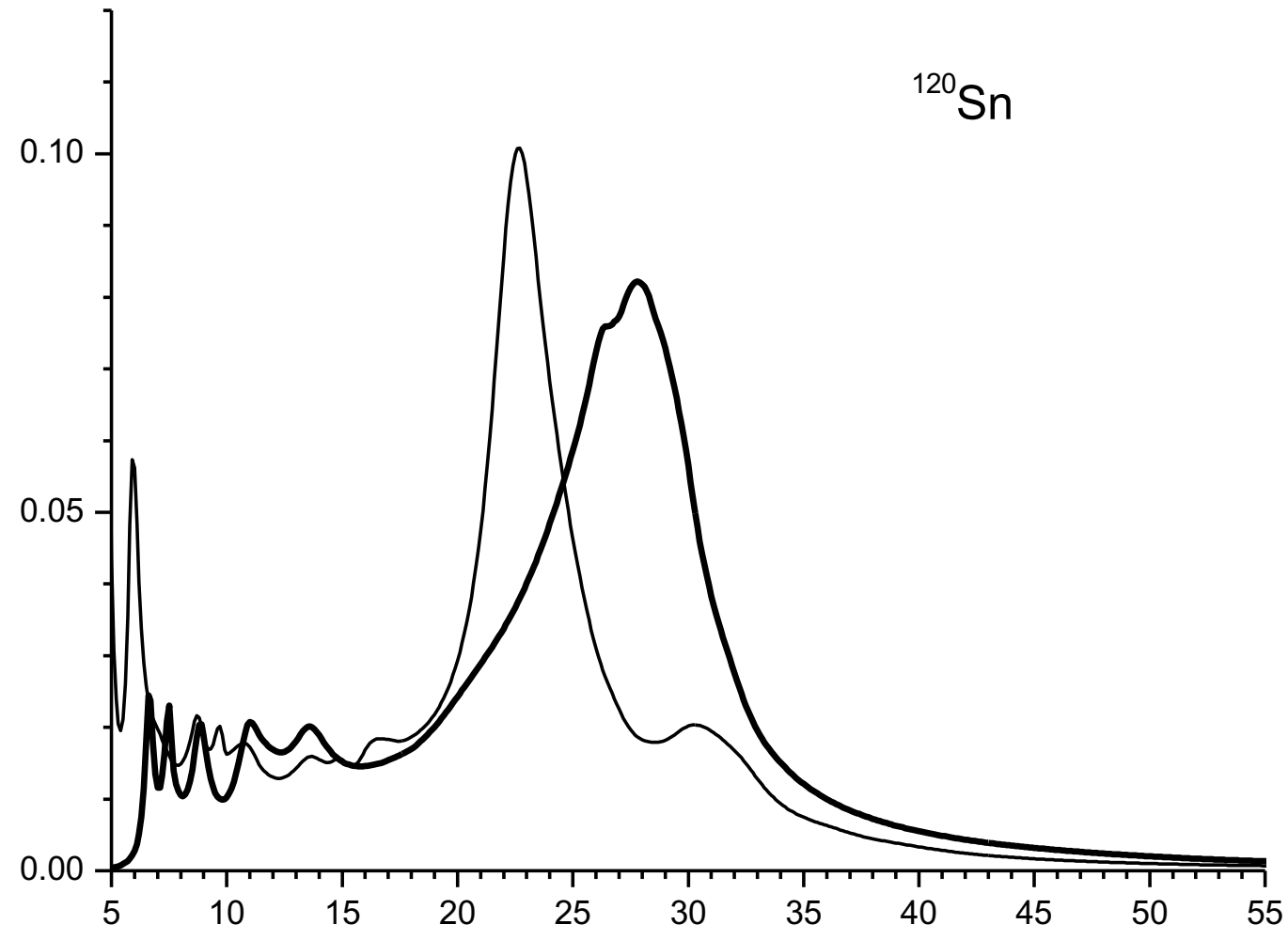

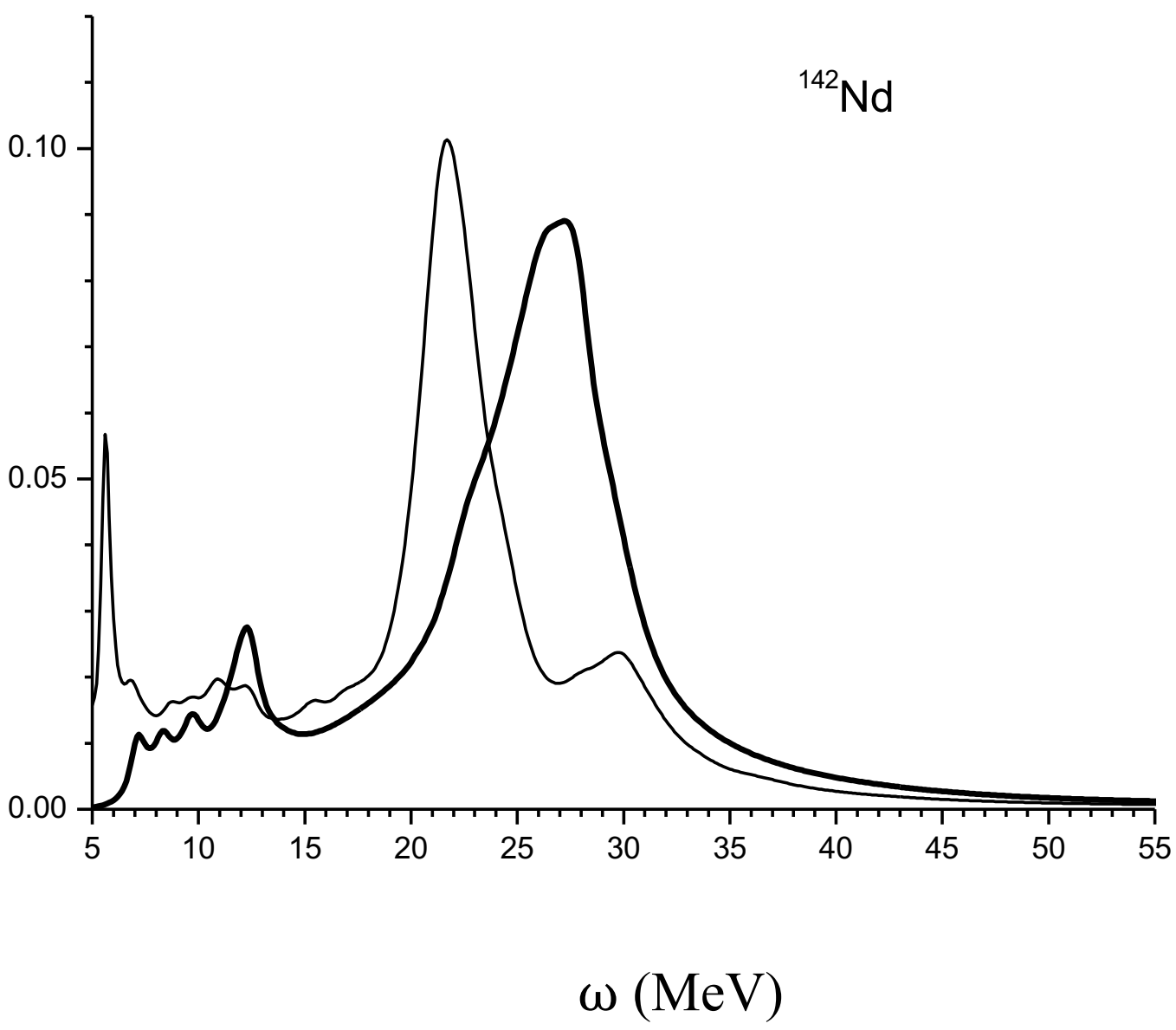

ω (MeV)

**FIG. 3.** The same as in Fig. 1 but for ISGDR (solid line) and ISGOR (thin line).

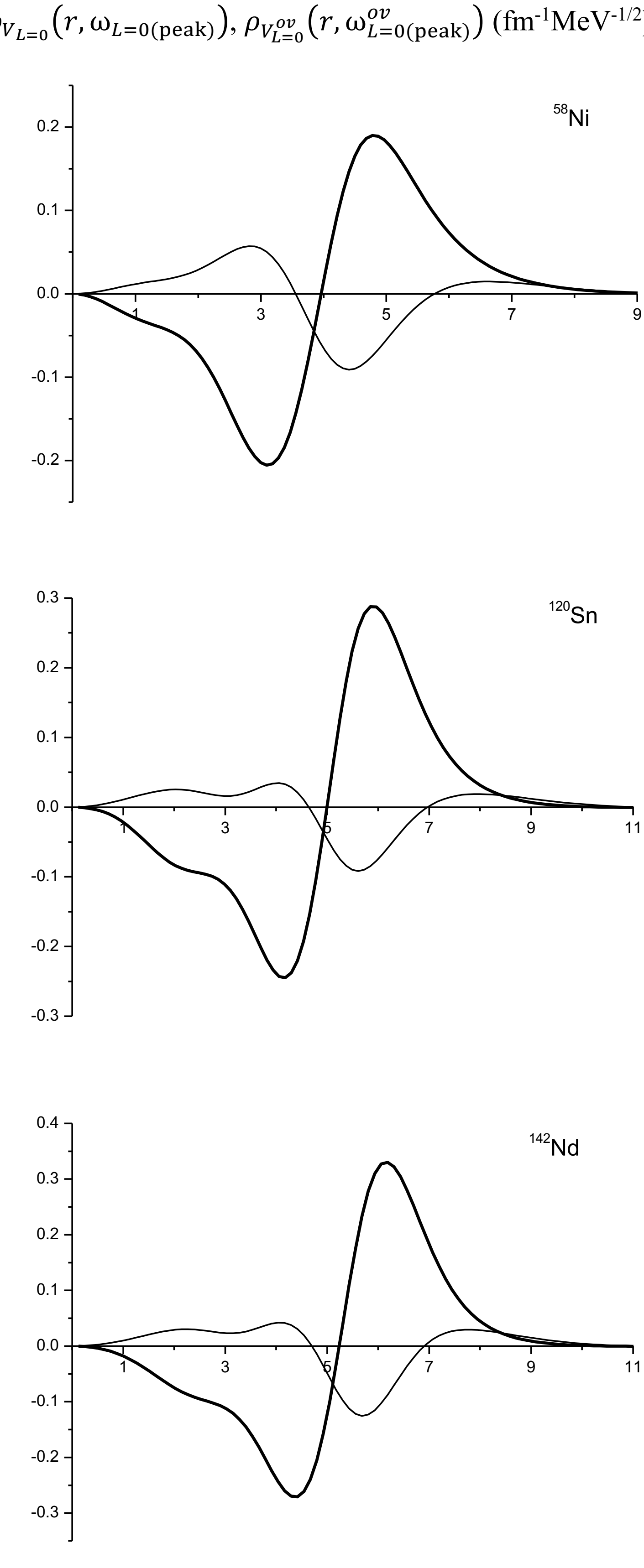

**FIG. 4.** The projected radial (one-dimensional) transition densities calculated within PHDOM and taken at the peak energy of ISGMR (solid line) and ISGMR2 (thin line) in nuclei under consideration.

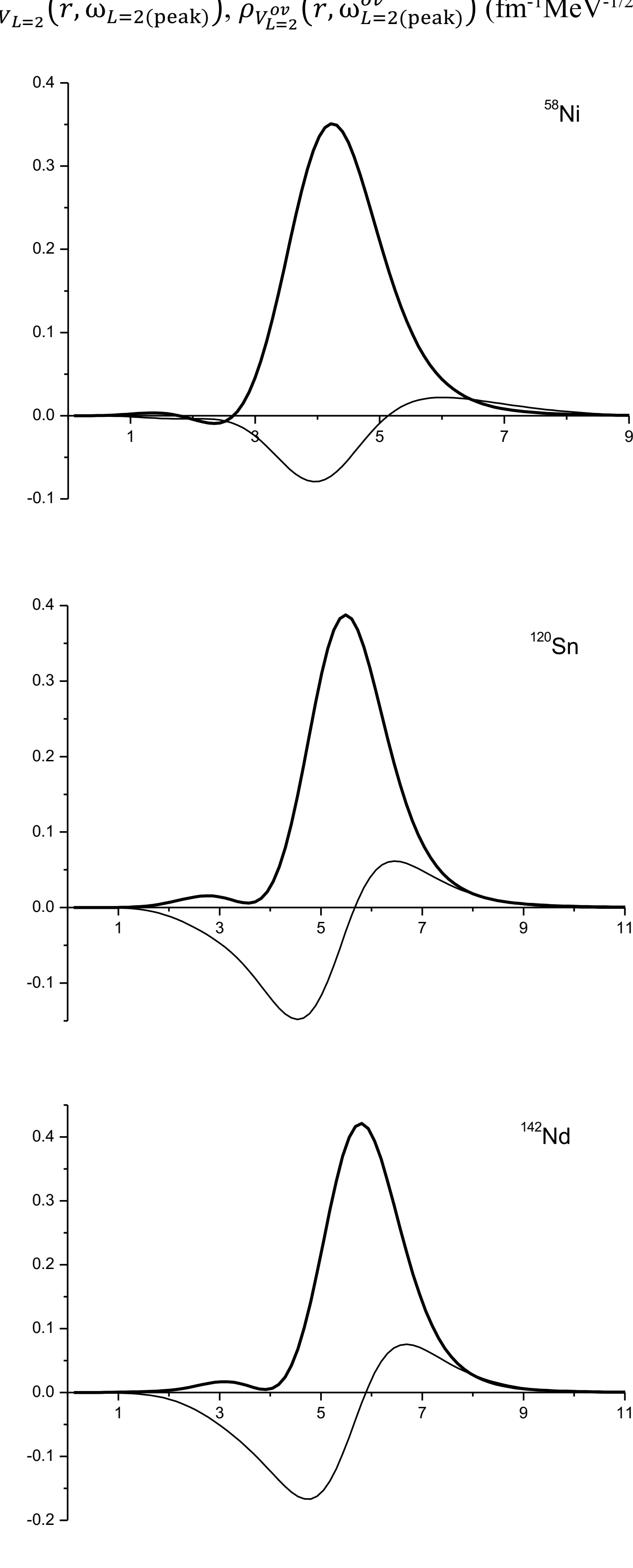

**FIG. 5.** The same as in Fig. 4 but for ISGQR (solid line) and ISGQR2 (thin line).

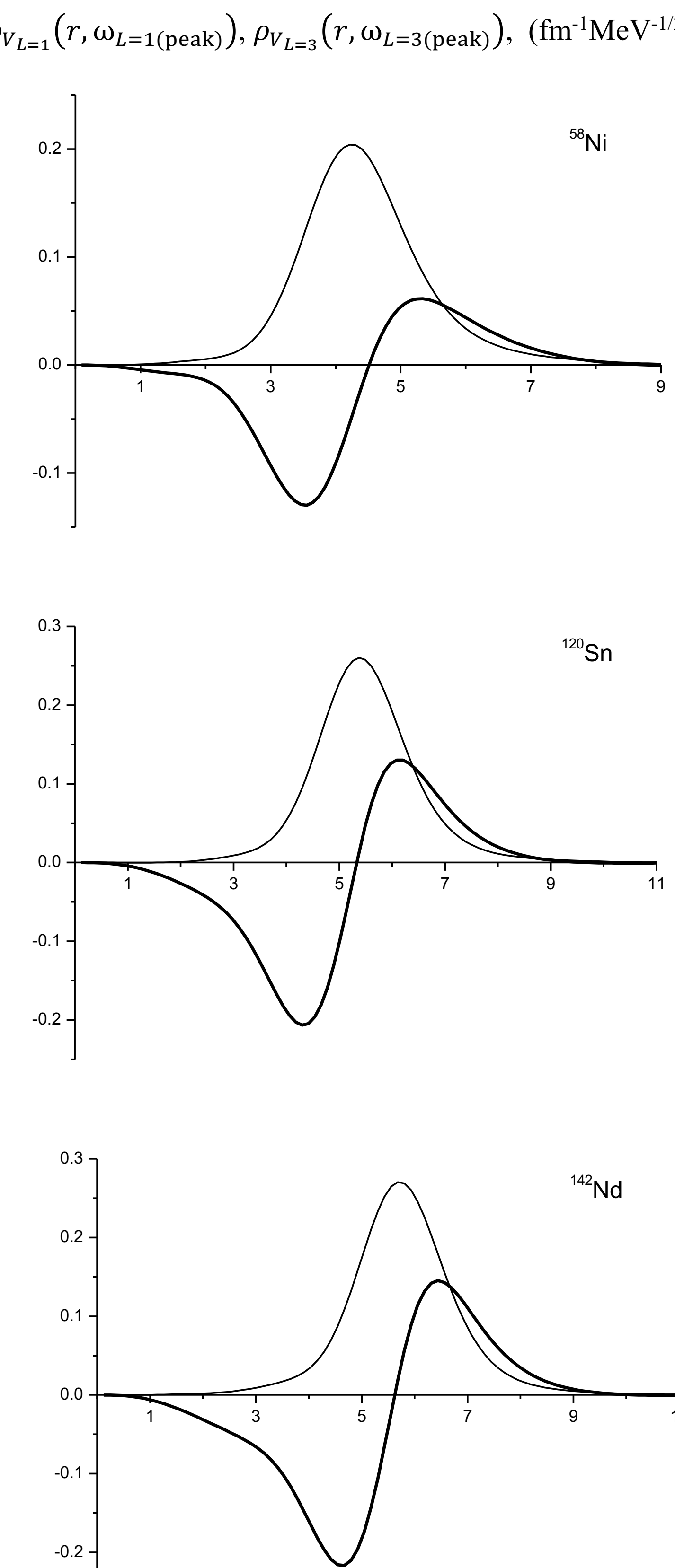

**FIG. 6.** The same as in Fig. 4 but for ISGDR (solid line) and ISGOR (thin line).

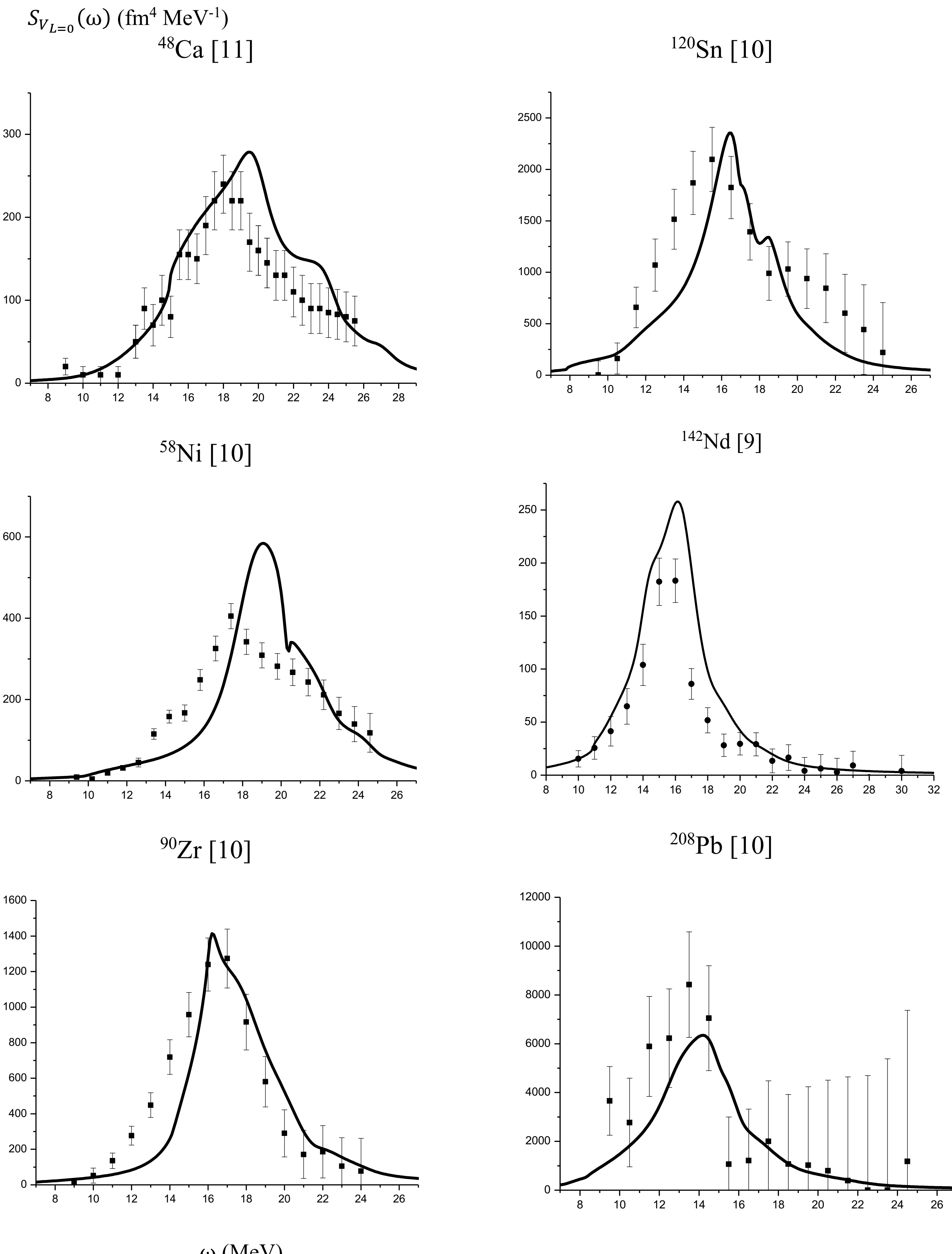

**FIG. 7.** A comparison of strength functions $S_{V_{L=0}}(\omega)$ evaluated within PHDOM for ISGMR in nuclei under consideration ($^{58}$Ni, $^{120}$Sn, $^{142}$Nd) and in nuclei $^{48}$Ca, $^{90}$Zr, $^{208}$Pb, considered previously (Refs. [5, 6]), with the respective experimental strength distributions (Refs. [9 - 11]).

$S_{V_{L=2}}(\omega)$ (fm$^4$ MeV$^{-1}$)

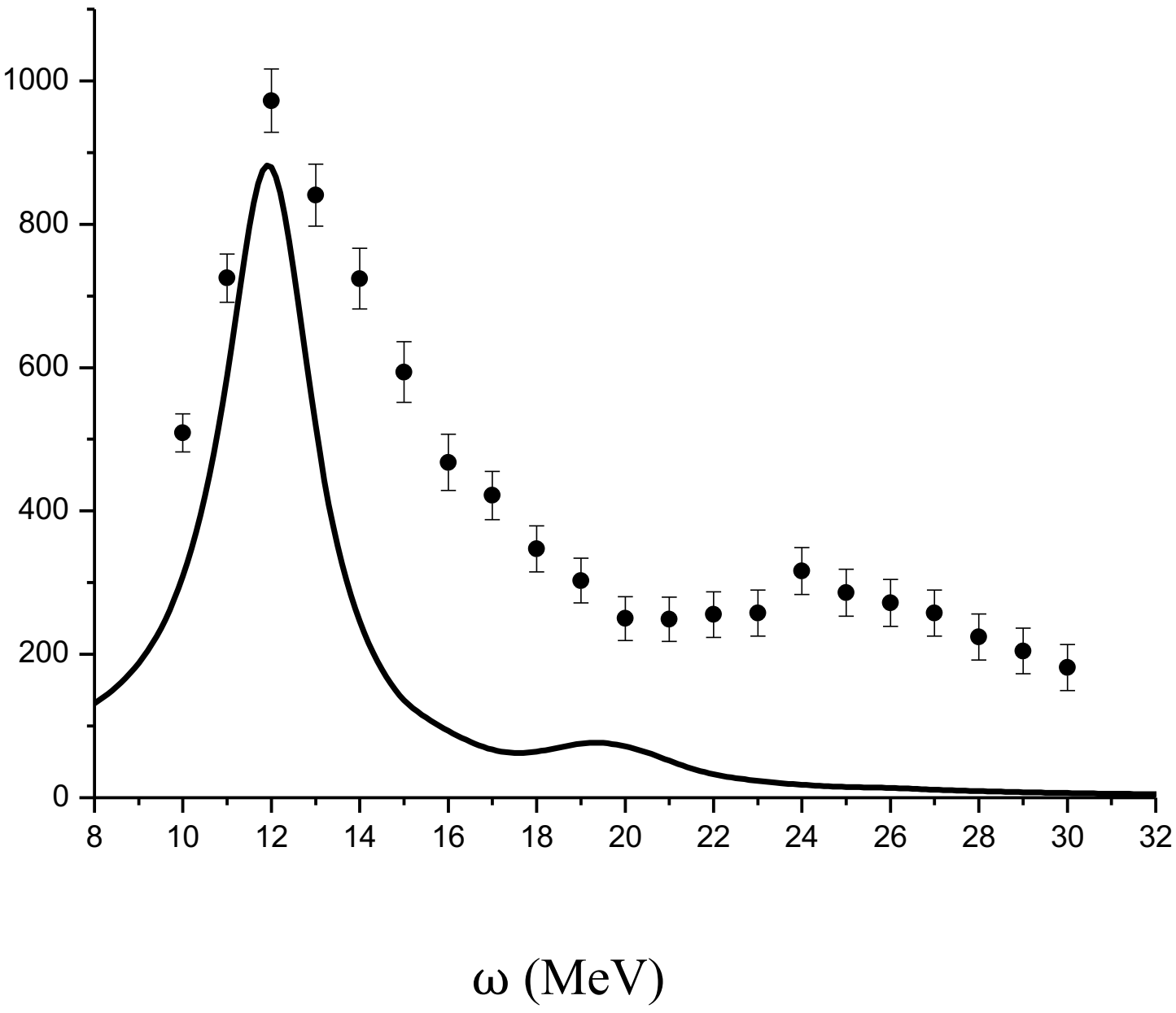

**FIG. 8.** A comparison of strength function $S_{V_{L=2}}(\omega)$ evaluated within PHDOM for ISGQR in $^{142}$Nd with the respective experimental strength distribution [9].

$S_{V_{L=2}^{ov}}(\omega)$ (fm$^8$ MeV$^{-1}$)

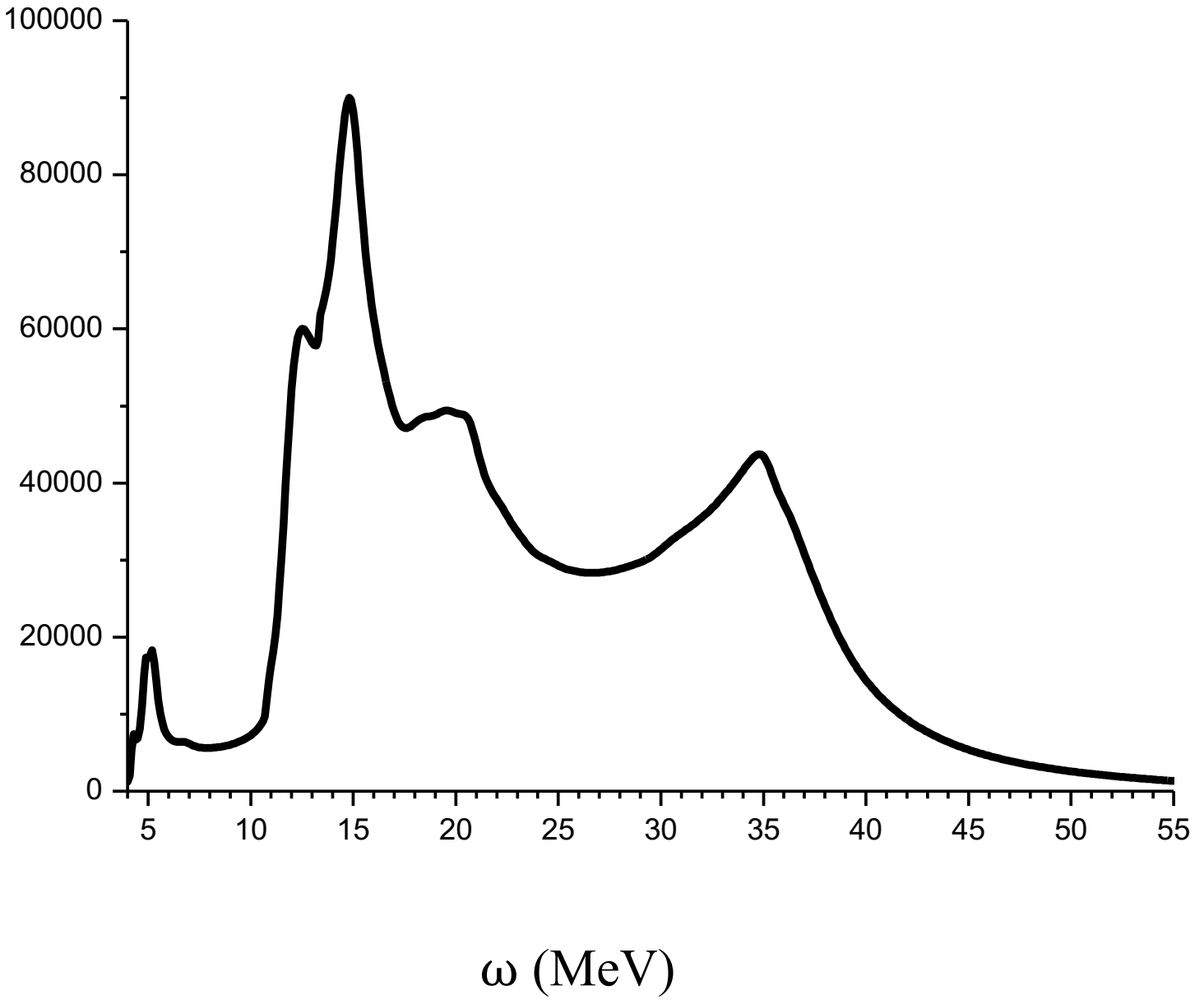

**FIG. 9.** The strength function $S_{V_{L=2}^{ov}}(\omega)$, evaluated within PHDOM for ISGQR2 in $^{142}$Nd.

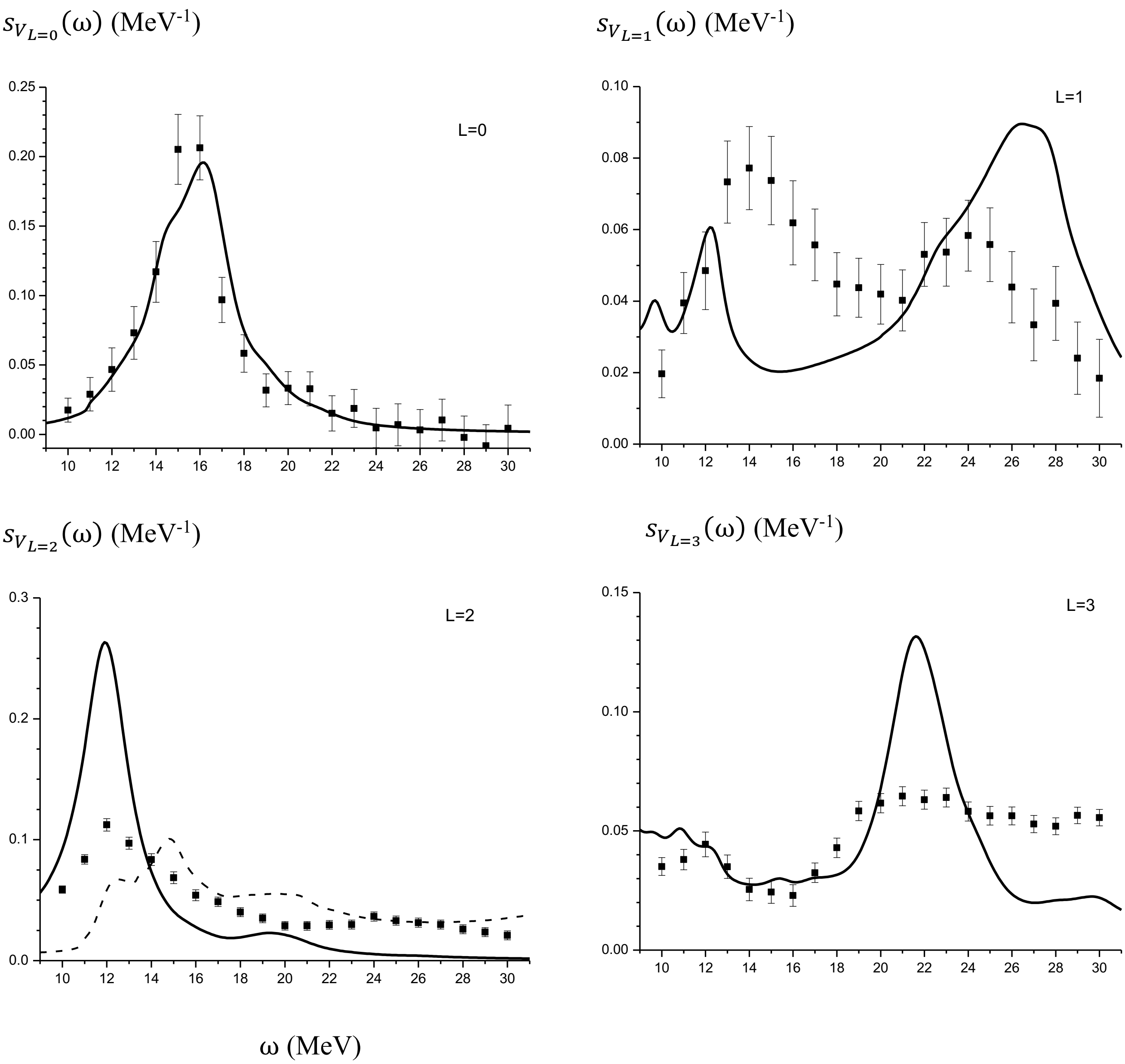

**FIG. 10.** A comparison of normalized strength functions $s_{V_L}(\omega)$ evaluated within PHDOM for ISGMPRs in $^{142}$Nd with the respective experimental normalized strength distributions [9] (the dashed line on the figure for $L = 2$ corresponds to low-energy part of the ISGQR2, $s_{V_{L=2}^{ov}}(\omega)$).